\documentclass[aps,showpacs,amsmath,amssymb,superscriptaddress,prl,preprint]{revtex4-1}
\usepackage{bm}
\usepackage{color}
\usepackage{epstopdf}
\usepackage[dvipdfmx]{graphicx}

\renewcommand{\thefootnote}{\fnsymbol{footnote}}

\begin{document}

\title{Intrinsic spin-orbit torque arising from Berry curvature in metallic-magnet/Cu-oxide interface}

\author{Tenghua Gao}
\affiliation{Department of Applied Physics and Physico-Informatics, Keio University, Yokohama 223-8522, Japan}

\author{Alireza Qaiumzadeh}
\affiliation{Center for Quantum Spintronics, Department of Physics, Norwegian University of Science and Technology, NO-7491 Trondheim, Norway}
\affiliation{Department of Physics, Institute for Advanced Studies in Basic Sciences (IASBS), Zanjan 45137-66731, Iran}

\author{Hongyu An}
\affiliation{Department of Applied Physics and Physico-Informatics, Keio University, Yokohama 223-8522, Japan}

\author{Akira Musha}
\affiliation{Department of Applied Physics and Physico-Informatics, Keio University, Yokohama 223-8522, Japan}

\author{Yuito Kageyama}
\affiliation{Department of Applied Physics and Physico-Informatics, Keio University, Yokohama 223-8522, Japan}

\author{Ji Shi}
\affiliation{School of Materials and Chemical Technology, Tokyo Institute of Technology, Tokyo 152-8552, Japan}

\author{Kazuya Ando\footnote{Correspondence and requests for materials should be addressed to ando@appi.keio.ac.jp}}
\affiliation{Department of Applied Physics and Physico-Informatics, Keio University, Yokohama 223-8522, Japan}

\begin{abstract}
We report the observation of the intrinsic damping-like spin-orbit torque (SOT) arising from the Berry curvature in metallic-magnet/CuO$_x$ heterostructures. We show that a robust damping-like SOT, an order of magnitude larger than a field-like SOT, is generated in the heterostructure despite the absence of the bulk spin-orbit effect in the CuO$_x$ layer.
Furthermore, by tuning the interfacial oxidation level, we demonstrate that the field-like SOT changes drastically and even switches its sign, which originates from oxygen modulated spin-dependent disorder. These results provide an important information for fundamental understanding of the physics of the SOTs.
\end{abstract}

\maketitle

The emergence of exciting field of spin-orbitronics~\cite{nphys2957,Kuschel} requires the fundamental understanding of spin-orbit torques (SOTs), which trigger magnetic dynamics via the exchange of angular momentum from carriers/crystal lattice to local magnetization~\cite{Berger,PhysRevB.66.014407,PhysRevB.72.033203,1038Kurebayashi}. The SOTs, 
damping-like (DL) and field-like (FL) torques, can arise from both bulk and interfacial relativistic spin-orbit interactions (SOIs).
In a ferromagnetic-metal/heavy-metal (FM/HM) heterostructure, a spin current is generated from spin-dependent scattering due to the bulk SOI in the HM, which is known as the spin Hall effect (SHE)~\cite{JETP,PhysRevLett.83.1834,science1087128,PhysRevLett.92.126603,science.1105514}. This spin current can exert a larger DL torque relative to a FL torque through the spin-transfer mechanism~\cite{Ralph20081190,nmat3311,nmat3823}. The other source for the SOTs is the Rashba-Edelstein effect due to the interfacial SOI~\cite{edelstein1990spin,Ganichev,sanchez2013spin,nmat4360}, which refers to the creation of nonequilibrium spin polarization at the HM/FM interface with broken inversion-symmetry. Although the Rashba-Edelstein effect primarily generates a large FL torque through spin exchange coupling, recent theoretical studies predict that a comparable DL torque in magnitude as a FL torque can be generated by taking into account spin-dependent scattering in a three-dimensional model of the interfacial SOI~\cite{PhysRevB.94.104420,PhysRevB.94.104419,PhysRevB.96.104438}. Moreover, theory and experiment demonstrate that the intrinsic mechanism of the SOT generation with the Berry curvature origin can produce a sizable DL component in a diluted magnetic semiconductor (DMS)~\cite{1038Kurebayashi,PhysRevB.91.134402}, and the existence of this intrinsic SOT is also expected in metallic bilayers, such as a Pt/Co bilayer~\cite{PhysRevB.91.134402,PhysRevB.92.014402,RevModPhys.87.1213,PhysRevB.90.174423}. Since the SOTs generated from all the contributions above have the same symmetry, it is a great experimental challenge to distinguish the mechanisms, consequently hindering the efficient engineering of the SOTs.

A promising system for studying the current-induced spin-orbit effect purely arising from the interfacial SOI is FM/insulating-oxide heterostructures, where the bulk spin-orbit effect can be neglected due to the insulating nature. Among the various oxides, Cu oxides (Cu$_2$O and CuO) have been intensively studied in a wide range of fields due to its abundant physical properties, such as ferromagnetism in ZnO based DMS~\cite{1.2032588,PhysRevB.74.075206,PhysRevLett.105.207201} and commensurate antiferromagnetic order at low temperature~\cite{PhysRevB.39.4343,science.1201061,PhysRevB.84.161405}. Furthermore, a recent study has demonstrated that Cu becomes an efficient SOT generator through oxidation, even though non-oxidized Cu possesses weak SOI~\cite{ncomms13069}. The efficient SOT generation, combined with the great flexibility of the oxidation level of Cu oxides, promises a way to study the physics of the SOTs purely generated by the interfacial SOI.

In this Letter, we demonstrate that the intrinsic Berry-curvature mechanism is responsible for the DL-SOT generation in Ni$_{81}$Fe$_{19}$/CuO$_x$ bilayers. In the bilayers where the CuO$_x$ layer is highly oxidized and semi-insulating, we observe a sizable DL-SOT in spite of the fact that the SOTs are purely generated by the interfacial SOI. We further found that the great flexibility of the oxidation level of Cu enables us to tune and even reverse the sign of the FL-SOT, opening a new avenue of SOT engineering. These features are consistent with the prediction of a two-dimensional (2D) Rashba model with oxygen modulated spin-dependent disorder. 

\begin{figure}[bt]
\includegraphics[scale=1]{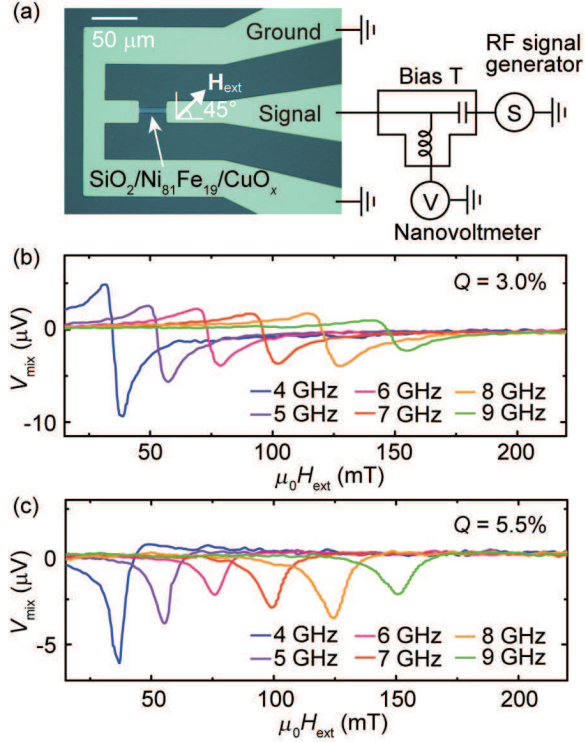}
\caption{
(a) An optical image of the sample geometry including contact pads, with the circuit and a $45^\circ$ tilt of an in-plane magnetic field ${\bf H}_\text{ext}$ with respect to the strip length direction used for the ST-FMR measurements. The $ H_\text{ext}$ dependence of the DC voltage $V_\text{mix}$ for the Ni$_{81}$Fe$_{19}$(7.5 nm)/CuO$_{x}$(10 nm) bilayers with (b) $Q=3.0$\% and (c) 5.5\% measured at the RF current frequencies of 4-9 GHz.}
\label{fig1}
\end{figure}

We used the spin-torque ferromagnetic resonance (ST-FMR)~\cite{PhysRevLett.106.036601,fang2011spin,Liu_science} to quantify the 
SOTs in the Ni$_{81}$Fe$_{19}$/CuO$_x$ bilayers at room temperature. The bilayers were fabricated by radio frequency (RF) magnetron sputtering in the following sequence. A 10-nm-thick CuO$_x$ layer was firstly grown on a thermally oxidized Si substrate by reactive sputtering, in a mixture of argon and oxygen atmosphere of 0.25 Pa. To manipulate the oxidation level of the CuO$_x$ layer, the oxygen to argon gas flow ratio ($Q$) was varied from 2.5\% to 5.5\%. Then, on the top of the semi-insulating CuO$_x$ layer, a Ni$_{81}$Fe$_{19}$ layer with the thickness ($t_\text{FM}$) of 7.5 nm was grown at argon pressure of 0.2 Pa, followed by a 4-nm-thick SiO$_2$ capping layer to prevent the oxidation of the Ni$_{81}$Fe$_{19}$ surface. The bilayers were patterned into rectangular strips with 4-$\mu$m width and 30-$\mu$m length by photolithography and liftoff techniques (see Fig.~\ref{fig1}(a)). For the ST-FMR measurement, an RF current with the frequency of $f$ was applied along the longitudinal direction of the device and an in-plane external field $H_\text{ext}$ was applied at an angle of $45^\circ$ with respect to the longitudinal direction of the device. The RF current generates the SOTs, which excite magnetic precession. The magnetization precession in the Ni$_{81}$Fe$_{19}$ layer causes the variation of the resistance owing to the anisotropic magnetoresistance (AMR). Therefore, the SOTs can be quantitatively determined by measuring a direct-current voltage, which is generated from the frequency mixing of the RF current and the oscillating resistance~\cite{Juretschke,10.1063/1.2400058,PhysRevB.78.104401}.

\begin{figure}[bt]
\includegraphics[scale=1]{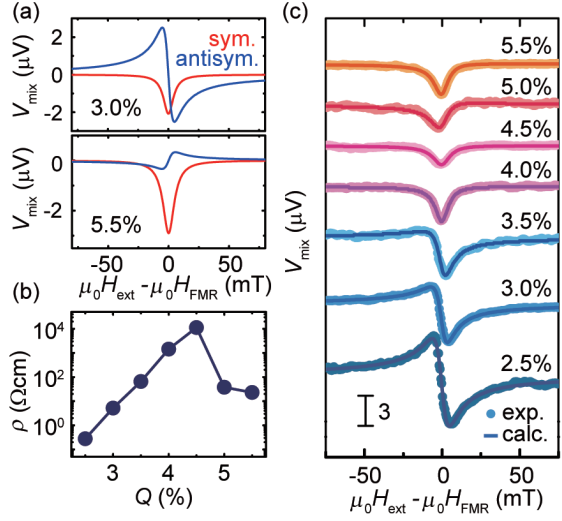}
\caption{
(a) The fitting curves of $V_\text{mix}$ as a function of the field for the Ni$_{81}$Fe$_{19}$(7.5 nm)/CuO$_{x}$(10 nm) bilayers with $Q=3.0$\% and 5.5\% at 7 GHz. The red and blue curves correspond to the symmetric and antisymmetric Lorentzian fitting, respectively. (b) The $Q$ dependence of the electrical resistivity $\rho$ of CuO$_{x}$(10 nm) single layer films, capped with 4 nm thick SiO$_2$ protective layer. The resistivity was measured by the four-probe method. (c) The $V_\text{mix}$ as a function of the field for the Ni$_{81}$Fe$_{19}$(7.5 nm)/CuO$_{x}$(10 nm) bilayers with various $Q$ measured at 7 GHz.}
\label{fig2}
\end{figure}

Figures~\ref{fig1}(b) and~\ref{fig1}(c) show the ST-FMR signals measured for the Ni$_{81}$Fe$_{19}$/CuO$_x$ bilayers with $Q=3.0$\% and 5.5\%, respectively.
The measured ST-FMR signals $V_\text{mix}$ can be expressed as the sum of symmetric and antisymmetric Lorentzian functions~\cite{PhysRevLett.106.036601,fang2011spin}: $V_\text{mix}= V_\text{s} L_\text{sym}(H_\text{ext})+V_\text{a}L_\text{asy}(H_\text{ext})$, where $L_\text{sym}(H_\text{ext})=W^2/[{(\mu_{0}H_\text{ext}-\mu_{0}H_\text{FMR})^2+W^2}]$ and $L_\text{asy}(H_\text{ext})=W(\mu_{0}H_\text{ext}-\mu_{0}H_\text{FMR})/[{(\mu_{0}H_\text{ext}-\mu_{0}H_\text{FMR})^2+W^2}]$. Here, $W$ and $\mu_0 H_\text{FMR}$ are the linewidth and the FMR field, respectively. Figure~\ref{fig2}(a) shows the symmetric and antisymmetric components of the measured ST-FMR signals, extracted by fitting the experimental data using $V_\text{mix}= V_\text{s} L_\text{sym}(H_\text{ext})+V_\text{a}L_\text{asy}(H_\text{ext})$. The extracted curves demonstrate that large $V_\text{s}$ signals are generated in both samples and the sign of $V_\text{a}$ is opposite in the devices with $Q=3.0$\% and 5.5\%.

The opposite sign of the antisymmetric component of the ST-FMR signals shows that the direction of the current-induced in-plane field in the Ni$_{81}$Fe$_{19}$/CuO$_x$ bilayers is reversed by changing the oxidation level of the CuO$_x$ layer. In ST-FMR signals, 
the symmetric component {$V$}$_\text{s}$ is proportional to the out-of-plane DL effective field $H_\text{DL}$, and the antisymmetric component {$V$}$_\text{a}$ corresponds to the in-plane field $H_\parallel$ due to the Oersted field $H_\text{Oe}$ and FL effective field $H_\text{FL}$: $H_\parallel=H_\text{Oe}+H_\text{FL}$~\cite{pai2015dependence}. To investigate the influence of the oxidation level on the SOT generation, we summarized $Q$ dependence of the resistivity of CuO$_x$ films in Fig.~\ref{fig2}(b) and the ST-FMR voltage of the Ni$_{81}$Fe$_{19}$/CuO$_x$ bilayers measured at 7 GHz in Fig.~\ref{fig2}(c). It can be clearly seen that the $V_\text s$ signal survives in the present entire $Q$ range, indicating the generation of the DL-SOT. On the other hand, the $V_\text a$ signal, or $H_\parallel$, decreases with raising the oxidation level, and even switches its sign at high $Q$ values. The sign reversal of $H_\parallel$ is also supported by the second harmonic Hall voltage measurements for the Ni$_{81}$Fe$_{19}$/CuO$_x$ bilayers (for details, see~\cite{Supplemental}).

\begin{figure}[bt]
\includegraphics[scale=1]{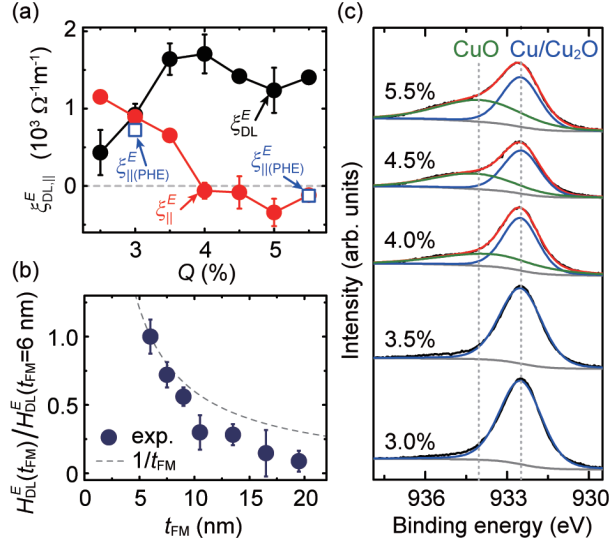}
\caption{
(a) The estimated SOT efficiency per unit electric field $\xi^{E}_{\text{DL}(\parallel)}$ for the Ni$_{81}$Fe$_{19}$(7.5 nm)/CuO$_x$(10 nm) bilayers with various $Q$ values. The open squares are the in-plane torque efficiency $\xi^{E}_{\parallel \text{(PHE)}}$ evaluated from the second harmonic Hall voltage measurements. The red and black solid circles are $\xi^{E}_\parallel$ and $\xi^{E}_\text{DL}$, respectively, estimated from the ST-FMR measurements. (b) The Ni$_{81}$Fe$_{19}$-layer-thickness $t_\text{FM}$ dependence of $H_\text{DL}^E=H_\text{DL}/E$ for the Ni$_{81}$Fe$_{19}$($t_\text{FM}$)/CuO$_x$(10 nm) bilayer at $Q=3$\%, where $H_\text{DL}$ is the DL spin-orbit effective field. The solid circles are the experimental data and the dashed curve is a function proportional to $1/t_\text{FM}$. (c) Curves fitting of Cu 2$p_{3/2}$ XPS spectra for CuO$_x$(10 nm) single layer films with various $Q$ values. The red fitting curves are the merged Cu/Cu$_2$O (blue curves) and CuO (green curves) 2$p_{3/2}$ peaks, and the grey curves are the Shirley background.}
\label{fig3}
\end{figure}

The DL spin-orbit effective field $H_\text{DL}$ and in-plane field $H_\parallel$ in the Ni$_{81}$Fe$_{19}$/CuO$_x$ bilayers can be quantified from the ST-FMR signals shown in Fig.~\ref{fig2}(c) using~\cite{1038Kurebayashi,fang2011spin,PhysRevB.92.214406}
\begin{equation}
V_\text{s}=\frac{I_\text{RF}\Delta R}{2}H_\text{DL}\frac{\gamma(\mu_{0}H_\text{FMR}+\mu_{0}M_\text{eff})\mu_{0}H_\text{FMR}}{2\sqrt{2} \pi f W(2\mu_{0}H_\text{FMR}+\mu_{0}M_\text{eff})},
\label{Heff1}
\end{equation}
\begin{equation}
V_\text{a}=\frac{I_\text{RF}\Delta R}{2}H_\parallel \frac{(\mu_{0}H_\text{FMR}+\mu_{0}M_\text{eff})}{\sqrt{2}W(2\mu_{0}H_\text{FMR}+\mu_{0}M_\text{eff})},
\label{Heff2}
\end{equation}
where $I_\text{RF}$ is the RF current in the strip (for details, see \cite{Supplemental}), $\Delta R$ is the AMR amplitude, $\gamma$ is the gyromagnetic ratio, and $\mu_0 M_\text{eff}$ is the demagnetization field.
In Fig.~\ref{fig3}(a), we tentatively estimate the $Q$ dependence of the torque efficiency per unit electric field $E$ for the thickness of the FM layer $t_\text{FM}= 7.5$ nm, defined by~\cite{nguyen2015spin}
\begin{equation}
\xi^{E}_{\text{DL}(\parallel)}=\frac{2e}{\hbar}\mu_{0}M_\text{s}t_\text{FM}\frac{H_{\text{DL}(\parallel)}}{E},
\label{xiFMR}
\end{equation}
where $M_\text{s}$ denotes the saturation magnetization. At the initial stage, $\xi^{E}_\text{DL}$ increases nearly four times before reaching the point of $Q$=3.5\%, and then becomes almost constant at the high oxidation level. One possible reason for the initial increase of $\xi^{E}_\text{DL}$ can be accounted for the enhanced Rashba parameter upon the formation of the oxide-metal interface, related to the effective electric field induced by the asymmetric charge distribution of the interface state, which is reminiscent of that observed at the Gd(0001) surface~\cite{PhysRevB.71.201403}. Similar enhancement of the DL-SOT has also been observed in W(O)/CoFeB~\cite{ncomms10644} and Pt/oxidized-CoFeB systems~\cite{nnano.2015.18}. In contrast to the increase of $\xi^{E}_\text{DL}$, $\xi^{E}_\parallel$ decreases monotonically with increasing the $Q$ values. The sign of $\xi^{E}_\parallel$ is reversed around $Q$=4\%, quantitatively consistent with $\xi^{E}_{\parallel\text{(PHE)}}$ obtained from the second harmonic Hall voltage measurement. 

The observed change of $\xi^{E}_\parallel$ originates from the sign reversal of the FL-SOT induced by changing the oxidation level at the Ni$_{81}$Fe$_{19}$/CuO$_x$ interface. It is worth noting that, in the Ni$_{81}$Fe$_{19}$/CuO$_x$ bilayers, the resistivity of the CuO$_x$ layers is more than three orders of magnitude higher than that of the Ni$_{81}$Fe$_{19}$ layer {($\sim$ 100 $\mu$$\Omega$cm)} as shown in Fig.~\ref{fig2}(b), indicating that $H_\text{Oe}$ due to current shunting through the CuO$_x$ layer is negligible in the Ni$_{81}$Fe$_{19}$/CuO$_x$ bilayers. 
However, $H_\text{Oe}$ can still be created by a possible nonuniform current distribution due to the different electron reflection at the top and bottom interfaces of the Ni$_{81}$Fe$_{19}$ layer~\cite{10.1063/1.3006005}. The contribution of $H_\text{Oe}$ to the observed $H_\parallel$ can be estimated by measuring the ST-FMR for 
a Ni$_{81}$Fe$_{19}$/CuO$_x$ film with thick Ni$_{81}$Fe$_{19}$. The reason for this is that the FL effective field $H_\text{FL}$ decreases with increasing the thickness $t_\text{FM}$ of the Ni$_{81}$Fe$_{19}$ layer, and we expect $H_\text{FL}\simeq 0$ and $H_\parallel \simeq H_\text{Oe}$ in the large $t_\text{FM}$ limit. From the ST-FMR for the thick film, we found $H_\text{Oe}>0$ for the Ni$_{81}$Fe$_{19}$/CuO$_x$ films with $Q=3$\% and 5.5\%. Using the Fuchs-Sondheimer model~\cite{10.1063/1.3006005} with the measured $H_\parallel$, we have confirmed $H_\text{FL}>0$ for the Ni$_{81}$Fe$_{19}$(7.5 nm)/CuO$_x$ film at $Q=3.0$\%. We note that at $Q>4$\%, $H_{\parallel}=H_\text{FL}+H_\text{Oe}<0$ as shown in Fig.~\ref{fig3}(a). This indicates $H_\text{FL}<0$ in the Ni$_{81}$Fe$_{19}$(7.5 nm)/CuO$_x$ films with higher $Q$. Thus, the sign of the FL-SOT is reversed from positive to negative by increasing $Q$ (for details, see Supplemental Materials~\cite{Supplemental})

The semi-insulating feature of the CuO$_x$ layer allows us to eliminate the generated DL-SOT from the spin-transfer mechanism of the SHE, since the charge current in the CuO$_x$ layer is negligible. Moreover, treating the interfacial SOI as a perturbation in ferromagnetic-metal/insulator bilayers, the imagery part of the interfacial SOT, or the DL-SOT, vanishes in a three-dimensional scenario regardless of any detail of a model~\cite{PhysRevB.96.104438}. This indicates that the extrinsic SOTs is unlikely to result in the efficient generation of the DL-SOT in the Ni$_{81}$Fe$_{19}$/CuO$_x$ bilayers. To further study the characteristic of the DL-SOT, we measured Ni$_{81}$Fe$_{19}$-layer-thickness $t_\text{FM}$ dependence of the DL effective field $H_\text{DL}/E$ for the Ni$_{81}$Fe$_{19}$($t_\text{FM}$)/CuO$_x$(10 nm) bilayer at $Q=3$\% as shown in Fig.~\ref{fig3}(b). The DL effective field decays faster than the $1/t_\text{FM}$ dependence, which is different from the $1/t_\text{FM}$ dependence of the SOT due to the bulk SHE~\cite{PhysRevB.87.174411,10.1038/ncomms4042}.

In the Ni$_{81}$Fe$_{19}$/CuO$_x$ bilayer, both the DL and FL-SOTs are generated by a SOI arising from the structural inversion symmetry breaking which is usually modeled by the Rashba SOI~\cite{PhysRevB.93.174421}. 
Since the carrier spins are exchange coupled to the magnetization in the Ni$_{81}$Fe$_{19}$ layer, the Ni$_{81}$Fe$_{19}$/CuO$_x$ bilayer can be approximately modeled as a 2D Rashba ferromagnet in which the itinerant spins are coupled to the localized spins via a \textit{sd} exchange interaction with a strength of $J_\text{ex}$. In this model, the Rashba-induced DL- and FL-SOTs are generated by two different scattering mechanisms: (i) the FL-SOT, $\bm{T}_\text{FL}\propto \bm{m} \times (\bm{z} \times \bm{E})$, originated from the scattering of spin carriers at the Fermi surface  with a conductivity like behavior, and (ii) the DL-SOT, $\bm{T}_\text{DL} \propto \bm{m} \times \bm{T}_\text{FL}$, with an intrinsic nature arising from the Berry phase curvature in the band structure; during the acceleration of carriers induced by the applied electric field, spins tilt and generate a non-equilibrium out-of-plane spin-polarization in response to an additional spin-orbit field, which gives rise to the intrinsic DL-SOT~\cite{1038Kurebayashi}.

In the strong exchange limit, microscopic calculations show that the FL-SOT is expressed as~\cite{PhysRevB.92.014402}
\begin{equation}
\bm{T}_\text{FL}\sim-2e\alpha_\text{R}\nu_\text{0}\left(\frac{\varepsilon_\text{F}+J_\text{ex}}{\gamma_{\uparrow}}-\frac{\varepsilon_\text{F}-J_\text{ex}}{\gamma_{\downarrow}}\right)
\bm{m} \times (\bm{z} \times \bm{E}),
\label{2Deq}
\end{equation}
where $\nu_{0}$, $\varepsilon_\text{F}$, and $\gamma_{\uparrow(\downarrow)}$ are the density of states per spin for a 2D electron gas, the Fermi energy, and the strength of the spin-dependent disorder scattering, respectively. Equation~(\ref{2Deq}) has three tunable parameters that can in principle explain the drastic change of the FL-SOT in the Ni$_{81}$Fe$_{19}$/CuO$_x$ bilayer with the oxidation: the variation of the SOI strength $\alpha_\text{R}$, the exchange strength $J_\text{ex}$, and the spin-dependent scattering rates $\gamma_{\uparrow(\downarrow)}$. First, the possible change of the Rashba SOI strength $\alpha_\text{R}$ at the interface induced by the oxidation cannot be responsible for the observed variation of the FL-SOT. According to the theory~\cite{PhysRevB.92.014402}, since both the DL- and FL-SOTs are linearly proportional to $\alpha_\text{R}$, they might have the same $Q$ dependence. This prediction is in sharp contrast to our observation shown in Fig.~\ref{fig3}(a).
Second, let us assume that the sign reversal of the FL-SOT is resulted from the sign reversal of $J_\text{ex}$ around $Q=4$\%. Under this assumption, $J_\text{ex}$ will be negligibly small around $Q=4$\%. On the other hand, in this region, i.e. the weak exchange limit, the theory predicts that the DL-SOT should be proportional to $J_\text{ex}^{2}$ while the FL-SOT is still linearly proportional to the exchange energy~\cite{PhysRevB.91.134402}. This indicates that an abrupt decrease of $\xi^{E}_\text{DL}$ should be observed around $Q=4$\%. This scenario also differs from our observation of a nearly constant $\xi^{E}_\text{DL}$ around $Q=4$\%, and thus the change of $J_\text{ex}$ is not significant in the Ni$_{81}$Fe$_{19}$/CuO$_x$ bilayer.

The origin of the observed sign change of the FL-SOT induced by the oxidation can be attributed to the variation of the spin-dependent disorder scattering. Assuming a metallic limit $\varepsilon_{F} \gg J_\text{ex}$, the term in the parentheses in Eq.~(\ref{2Deq}) can be simplified $\sim$ $\varepsilon_\text{F}(1/\gamma_{\uparrow}-1/\gamma_{\downarrow})$.
If the relative strength of the spin-dependent disorder scattering could be tuned through varying the interfacial oxidation level, it is possible to observe the sign reversal of the FL-SOT without changing the sign of the DL-SOT. The reason for this behavior originates from different scattering dependence of the two components of the Rashba SOTs; the FL-SOT has conductivity-like behavior and is sensitive to the spin-dependent scattering while the intrinsic DL-SOT is robust against disorders in the weak disorder regime. The sign change of the interfacial FL-SOT through tuning disorders was also predicted by $ab$ $initio$ calculations for more realistic band structures~\cite{PhysRevB.90.174423}. For permalloy, it is demonstrated that the minority spin states of Ni at the Fermi level is heavily damped by Fe impurities due to the greatly differed potentials for the two constituents~\cite{PhysRevB.65.075106}. Therefore, a change of $\gamma_{\downarrow}$ can be certainly expected if the concentration of interfacial permalloy is modulated by the interfacial oxidation level. 
We note that although $Q$ was varied only slightly, from 2.5\% to 5.5\%, the oxidation level of the CuO$_x$ layer is significantly changed, as evidenced in the drastic change of the resistivity $\rho$ (see Fig.~\ref{fig2}(b)). As shown in Fig.~\ref{fig2}(b), when the value of $Q$ increases from 2.5 to 5.5\%, the resistivity $\rho$ of the CuO$_x$ film initially increases, after approaching its highest value around $Q$=4.5\%, then reduced. This extraordinary tendency is because of the formation of various types of CuO$_x$, such as Cu$_2$O, CuO, or their mixture, most likely attributed to the stoichiometry related Cu vacancies~\cite{pssb.201248128}. This drastic change of the oxidation state of the CuO$_x$ layer can influence the oxidation level near the Ni$_{81}$Fe$_{19}$/CuO$_x$ interface. To further obtain the information on the oxidation at the interface, the X-ray photoelectron spectra (XPS) measurements were performed on CuO$_x$(10 nm) single layer films with various $Q$. As shown in Fig.~\ref{fig3}(c), the CuO phase appears around $Q=4$\%, which coincides with the oxidation level where the sign reversal of the FL-SOT is observed.

In conclusion, we demonstrated that the robust intrinsic DL-SOT with interfacial feature is generated in the Ni$_{81}$Fe$_{19}$/CuO$_x$ bilayers. 
Although the oxidation effect on the SOT generation in metallic heterostructures has been reported previously~\cite{nnano.2015.18}, the presence of a heavy metal layer makes it difficult to provide a physical picture of the SOT generation. In contrast, the semi-insulating feature of CuO$_x$ enables to reveal the physics behind the oxidation effect on the SOT generation. We noticed that the observed SOTs purely originate from the interfacial SOI in the Ni$_{81}$Fe$_{19}$/CuO$_x$  bilayers with different scattering mechanisms, i.e. the conductivity-like FL-SOT and the intrinsic DL-SOT, providing a basic understanding on the SOTs generation. Therefore, we believe that the spin-orbit device based on Cu oxide is an ideal system for the study of the intrinsic DL-SOT, as well as the interfacial oxidation-tuning SOTs. 

\begin{acknowledgments}
We thank Takashi Harumoto for assistance on the AES measurements. This work was supported by JSPS KAKENHI Grant Numbers 26220604, 26103004, the Asahi Glass Foundation, JSPS Core-to-Core Program, and Spintronics Research Network of Japan (Spin-RNJ). H.A. acknowledges the support from the JSPS Fellowship (No.~P17066).
A. Q. was supported by the European Research Council via Advanced Grant number 669442 ``Insulatronics", and the Research Council of Norway through
its Centres of Excellence funding scheme, project number 262633 ``QuSpin". We also thank the referee who helped, with his/her useful comments, to clarify the Oersted field contribution in the ST-FMR measurements.
\end{acknowledgments}

\clearpage


\renewcommand{\thefootnote}{\fnsymbol{footnote}}
\setcounter{figure}{0} 
\renewcommand{\thefigure}{S\arabic{figure}}
\renewcommand{\bibnumfmt}[1]{[S#1]}
\renewcommand{\citenumfont}[1]{S#1}

\begin{center}\textbf{Suplemental Materials for}

\vspace{6pt}

\textbf{\large{Intrinsic spin-orbit torque arising from Berry curvature \\in metallic-magnet/Cu-oxide interface}}

\vspace{12pt}

{Tenghua Gao, Alireza Qaiumzadeh, Hongyu An, Akira Musha, Yuito Kageyama, Ji Shi, and Kazuya Ando}

\end{center}

\vspace{12pt}

\noindent 
\textbf{1. Determination of in-plane effective field by second harmonic Hall voltage measurements}
 \vspace{12pt}

To verify the sign reversal of the in-plane effective field ($H_{\parallel}$), the second harmonic voltage measurements were preformed on the Ni$_{81}$Fe$_{19}$(7.5 nm)/CuO$_x$(10 nm) bilayers with $Q =3.0$\% and 5.5\% at room temperature~\cite{PhysRevB.90.224427.1,acs.nanolett.6b03300.1,PhysRevB.95.104403.1}. The Ni$_{81}$Fe$_{19}$/CuO$_x$ bilayers were patterned into Hall bars with dimensions of $L(70\:\mu\text{m})\times W(20\: \mu\text{m})$ by photolithography, and fabricated in the same batch as the corresponding mircostrip used for the spin-torque ferromagnetic resonance (ST-FMR) measurements. 

\begin{figure}[htb]
\includegraphics[scale=1]{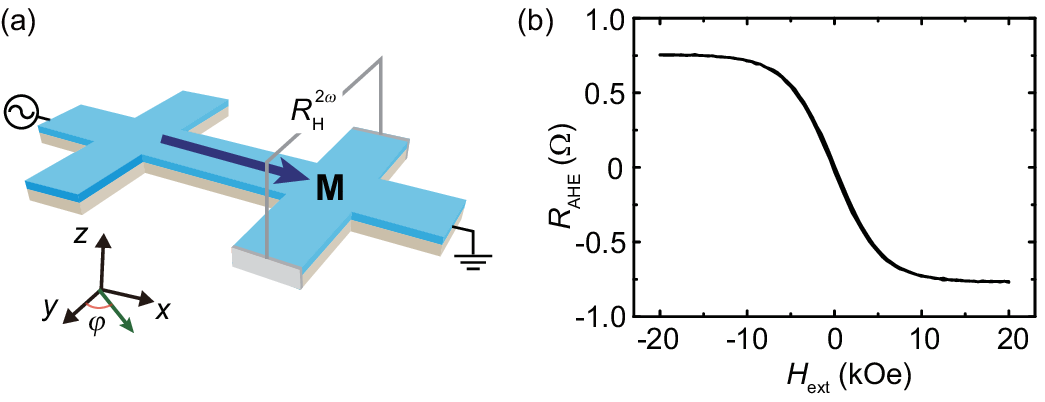}
\caption{(a) Schematic illustration of the second harmonic Hall voltage measurement for the Ni$_{81}$Fe$_{19}$(7.5 nm)/CuO$_x$(10 nm) Hall bars. (b) Anomalous Hall resistance $R_\text{AHE}$ as a function of $H_\text{ext}$ for the Ni$_{81}$Fe$_{19}$(7.5 nm)/CuO$_x$(10 nm) bilayers with $Q=3.0$\%.}
\label{figS1} 
\end{figure}

In general, the transverse resistance, namely the Hall resistance $R_\text{H}$, is composed of the contributions from anomalous Hall effect (AHE) $R_\text A$ and planar Hall effect (PHE) $R_\text P$ as
\begin{equation}
R_\text{H}=R_\text{A}\cos\theta_{m}+R_\text{P}\sin^{2}\theta_{m} \sin2\varphi_{m}
\tag{S1}\label{S1} ,
\end{equation} where $\theta_m$ and $\varphi_m$ are the polar and in-plane azimuthal angles for the magnetization, respectively. For the measurement of the Hall resistance, we applied an ac voltage of 5 V with a frequency of 35.85 Hz along the longitudinal direction of the Hall bars as shown in Fig.~\ref{figS1}(a). As a result, the magnetization precession is induced in the same frequency, oscillating with small angles of $\Delta \theta$ and $\Delta \varphi$. Since the Ni$_{81}$Fe$_{19}$(7.5 nm)/CuO$_x$(10 nm) bilayers have easy plane anisotropy with in-plane isotropy (feature of polycrystal) as confirmed in Fig.~\ref{figS1}(b), the magnetization almost lies in the film plane. This indicates that the demagnetizing field $\mid H_K\mid$ is much larger than the damping-like (DL) effective field $\mid H_\text{DL}\mid$. Consequently, the induced magnetization precession gives rise to the second harmonic Hall resistance $R^{2\omega}_\text{H}$~\cite{PhysRevB.90.224427.1,acs.nanolett.6b03300.1,PhysRevB.95.104403.1}:
\begin{equation}
R^{2\omega}_\text{H}=\left (\frac{1}{2}\Delta\theta R_\text{A}+\frac{V_\text{0}}{R}\alpha \Delta T\right )\cos\varphi_{m}-\Delta \varphi R_\text{P}\cos2\varphi_{m} \cos\varphi_{m}
\tag{S2}, \label{S2}
\end{equation} 
where the second term in the parentheses presents the contribution from the anomalous Nernst effect (ANE). $V_{0}$, $R$, $\alpha$, and  $\Delta T$ are the peak applied voltage, the longitudinal resistance of the Hall bar, the ANE coefficient, and the temperature gradient between the top and bottom layers, respectively. In the case that $\Delta \theta$ and $\Delta \varphi$ are sufficiently small, the deviating angles can be simplified to $\Delta \theta=H_\text{DL}/(-H_\text{K}+\mid H_\text{ext}\mid$), $\Delta \varphi=H_{\parallel}/\mid H_\text{ext}\mid$. Substituting into Eq.~(\ref{S2}) and defining $\varphi=\pi /2-\varphi_{m}$, we obtain
\begin{equation}
R^{2\omega}_\text{H}=\left (\frac{1}{2}\frac{H_\text{DL}}{H_\text{K}-\mid H_\text{ext}\mid}R_\text{A}+\frac{V_\text{0}}{R}\alpha \Delta T\right )\sin\varphi+ \frac{H_{\parallel}}{\mid H_\text{ext}\mid} R_\text{P}\cos2\varphi \sin\varphi.
\tag{S3}
\label{S3} 
\end{equation}

\begin{figure}[tb]
\includegraphics[scale=1]{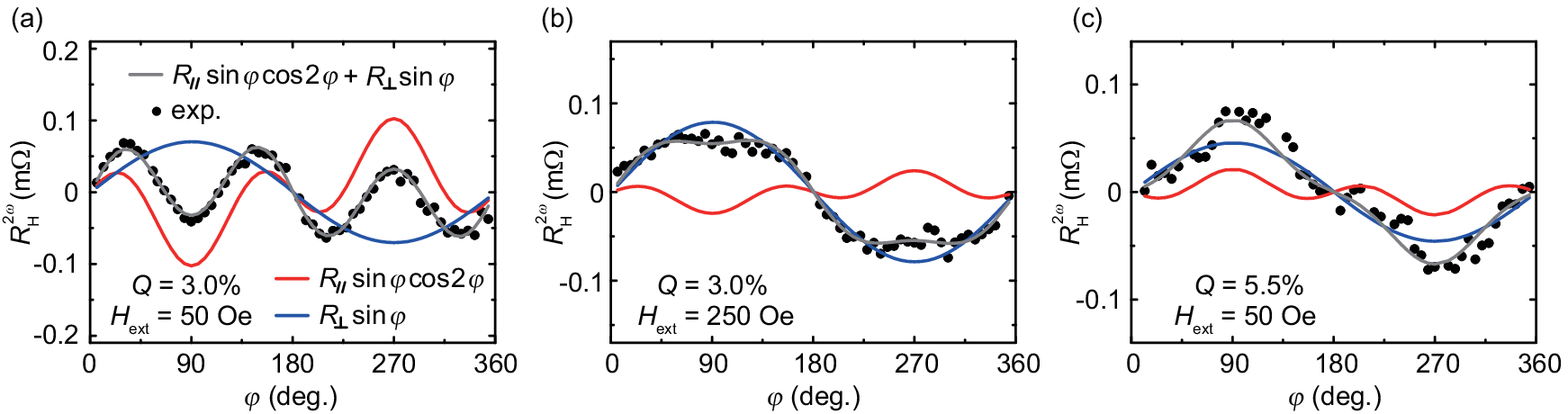}
\caption{The angular dependence of the second harmonic Hall resistance $R^{2\omega}_\text{H}$ measured for the Ni$_{81}$Fe$_{19}$(7.5 nm)/CuO$_x$(10 nm) bilayers with $Q=3.0\%$ at (a) $H_\text{ext}=50$ Oe and (b) $H_\text{ext} =250$ Oe. (c) The angular dependence of the second harmonic Hall resistance $R^{2\omega}_\text{H}$ measured for the Ni$_{81}$Fe$_{19}$(7.5 nm)/CuO$_x$(10 nm) bilayers with $Q=5.5\%$ at $H_\text{ext} =50$ Oe. The gray curve shows the fitting results using Eq.~(\ref{S3}). The red and blue curves correspond to the $\cos2\varphi \sin\varphi$ and $\sin\varphi$ components, respectively.}
\label{figS2} 
\end{figure}

Figures~\ref{figS2}(a) and \ref{figS2}(b) show the angular dependence of $R^{2\omega}_\text{H}$ for the Ni$_{81}$Fe$_{19}$(7.5 nm)/CuO$_x$(10 nm) bilayers with $Q$=3.0\% at $H_\text{ext}=50$ and 250 Oe, respectively. The second harmonic signals were recorded utilizing LI5640 lock in amplifier. Through fitting the data using Eq.~(\ref{S3}), the angular dependence of $R^{2\omega}_\text{H}$ can be divided into $\sin\varphi$ and $\cos2\varphi \sin\varphi$ components. When $H_\text{ext}$ increases from 50 to 250 Oe, the $\cos2\varphi \sin\varphi$ component, which corresponds to the in-plane effective field, is reduced, consistent with the $1/\mid H_\text{ext}\mid$ dependence. For the $\sin\varphi$ component, we note that $\mid H_\text{K}\mid$ obtained from AHE measurements is $\sim$ 1 T, which should be much larger than $\mid H_\text{DL}\mid$. Therefore, we can conclude that in the small $H_\text{ext}$ region, the ANE term dominates in the $\sin\varphi$ component. Regarding to the ANE, the thermal conductivities of both Cu$_2$O ($\kappa=5.6$ Wm$^{-1}$K$^{-1}$) and CuO ($\kappa=33$ Wm$^{-1}$K$^{-1}$) are larger than that of SiO$_2$ ($\kappa=1.4$ Wm$^{-1}$K$^{-1}$) protective layer, inducing a positive thermal gradient in the bilayers. Thus, the $\sin\varphi$ components of the Ni$_{81}$Fe$_{19}$/CuO$_x$ bilayers with $Q=3.0$\% and 5.5$\%$ have the same positive sign, although the oxidation level of the CuO$_x$ layer is different between these two cases. Comparing $R^{2\omega}_\text{H}$ for the Ni$_{81}$Fe$_{19}$/CuO$_x$ bilayers with $Q=3.0$\% and 5.5$\%$ shown in Figs.~\ref{figS2}(a)and \ref{figS2}(c), an unambiguous sign reversal of the $\cos2\varphi \sin\varphi$ component is observed. 

\begin{figure}[bt]
\includegraphics[scale=1]{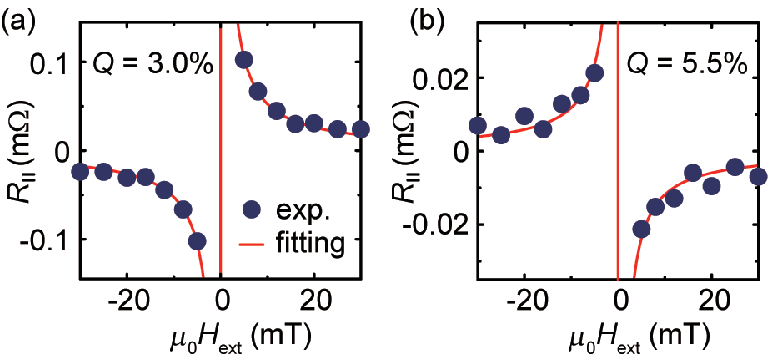}
\caption{
The second harmonic Hall voltage measurements for the Ni$_{81}$Fe$_{19}$(7.5 nm)/CuO$_x$(10 nm) bilayers with (a) $Q=3$\%, and (b) $Q=5.5$\%. The data points correspond to the extracted in-plane oscillating resistance $R_{\parallel}$ as a function of applied external magnetic field $H_\text{ext}$. The red curves are fits to $R_\text{P}H_\parallel/H_\text{ext}$.}
\label{fig2-2}
\end{figure}

Then, we extracted the $1/H_\text{ext}$ dependence of in-plane oscillating resistance $R_{\parallel}$ from these angular dependence measurements. Figures~\ref{fig2-2}(a) and~\ref{fig2-2}(b) show the $H_\text{ext}$ dependence of the extracted $R_{\parallel}$ for the Ni$_{81}$Fe$_{19}$/CuO$_x$ bilayers with $Q=3.0$\% and 5.5\%, respectively. In both two Ni$_{81}$Fe$_{19}$/CuO$_x$ bilayers with different $Q$, $R_{\parallel}$ changes its sign around $H_\text{ext}=0$ and becomes smaller with increasing $H_\text{ext}$. Here, $R_{\parallel}$ is related to the in-plane effective field $H_\parallel$ as $R_{\parallel}=R_\text{P}H_\parallel/H_\text{ext}$~\cite{acs.nanolett.6b03300.1}.  
Thus, the different sign of $R_{\parallel}$ shown in Figs.~\ref{fig2-2}(a) and \ref{fig2-2}(b) demonstrates that the sign of the in-plane effective field $H_\parallel$ is opposite between the Ni$_{81}$Fe$_{19}$/CuO$_x$ bilayers with $Q=3.0$\% and 5.5\%, which is consistent with the ST-FMR results.

\clearpage
\noindent  \textbf{2. Calibration of radio frequency current}
 \vspace{12pt}
 
\begin{figure}[hb]
\includegraphics[scale=1]{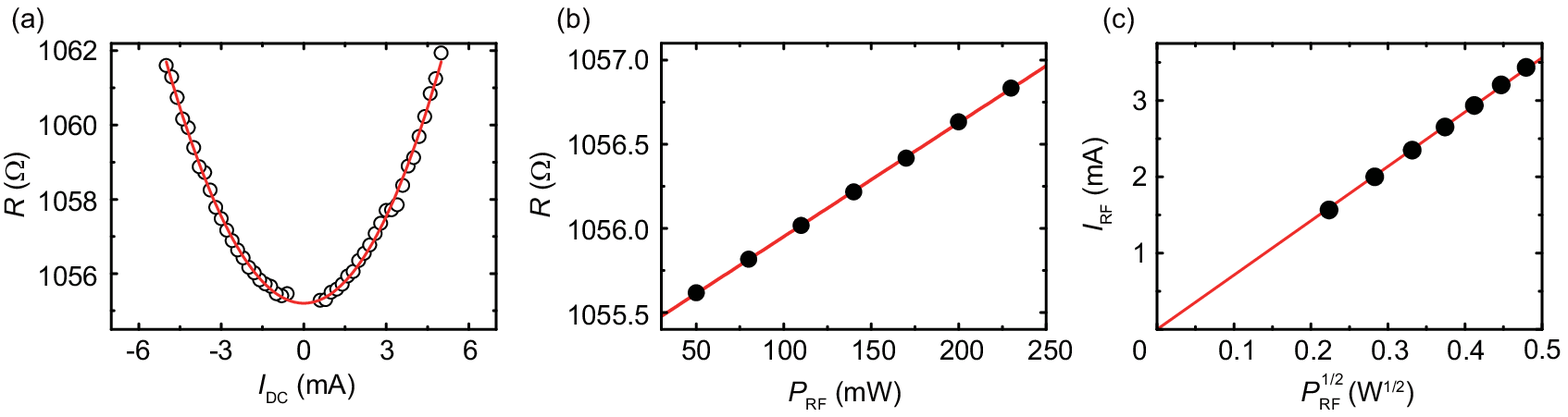}
\caption{Resistance $R$ change of the Ni$_{81}$Fe$_{19}$(7.5 nm)/CuO$_x$(10 nm) bilayer strip with $Q=5.5\%$ due to the Joule heating caused by the application of (a) DC current $I_\text{DC}$, where the red curve is the fitting result using a parabolic function, and (b) RF current, where $P_\text{RF}$ is the applied RF power. (c) The calibrated RF current $I_\text{RF}$ as a function of the square root of the applied RF power $P_\text{RF}$. The red line is the linear fitting to the data.}
\label{figS3} 
\end{figure}

For the ST-FMR measurements, the applied radio frequency (RF) power was 170 mW. The impedance mismatch between the microstripe and RF source results in the reflection of the applied RF current. To estimate the RF current flowing in the ST-FMR device, we made use of the current induced resistance change due to Joule heating~\cite{fang2011spin.1,PhysRevB.92.214406.1}. One example is shown in Fig.~\ref{figS3} for the Ni$_{81}$Fe$_{19}$(7.5 nm)/CuO$_x$(10 nm) bilayer with $Q=5.5\%$ and the size of $L(30 \:\mu\text{m})\times W(4 \:\mu\text{m}$). Figure~\ref{figS3}(a) shows the resulting resistance changes caused by the Joule heating due to DC current $I_\text{DC}$ application. The resistance change follows the parabolic relationship to the applied current, as expected for the sample heating. Then, we measured the resistance change due to the RF power $P_\text{RF}$ application, as shown in Fig.~\ref{figS3}(b). By comparing these changes due to the DC current and RF power applications, we can estimate the energy dissipation resulted from each applied RF power, as well as the corresponding RF current in the strip. Figure~\ref{figS3}(c) shows the calibrated RF current $I_\text{RF}$ flowing in the ST-FMR device as a function of the square root of the applied RF power $\sqrt{P_\text{RF}}$, where the corresponding RF current $I_\text{RF}$ is $\sqrt{2}$ times the DC current $I_\text{DC}$ because
the heating for the RF current is given by $I_\text{RF}^2R/2$ compared to $I_\text{DC}^2R$ for the DC current. It is clearly seen that as expected, $I_\text{RF}$ increases linearly with $\sqrt{P_\text{RF}}$. Therefore, we can extract the RF current for each applied power from the linear fitting. Following this method, the RF current for the Ni$_{81}$Fe$_{19}$(7.5 nm)/CuO$_x$(10 nm) bilayers with different $Q$ values was calibrated individually.

\clearpage

\noindent  \textbf{3. Auger electron spectroscopy depth profile analysis of Ni$_{81}$Fe$_{19}$/CuO$_x$ bilayers}
\vspace{12pt}

\begin{figure}[hb]
\includegraphics[scale=1]{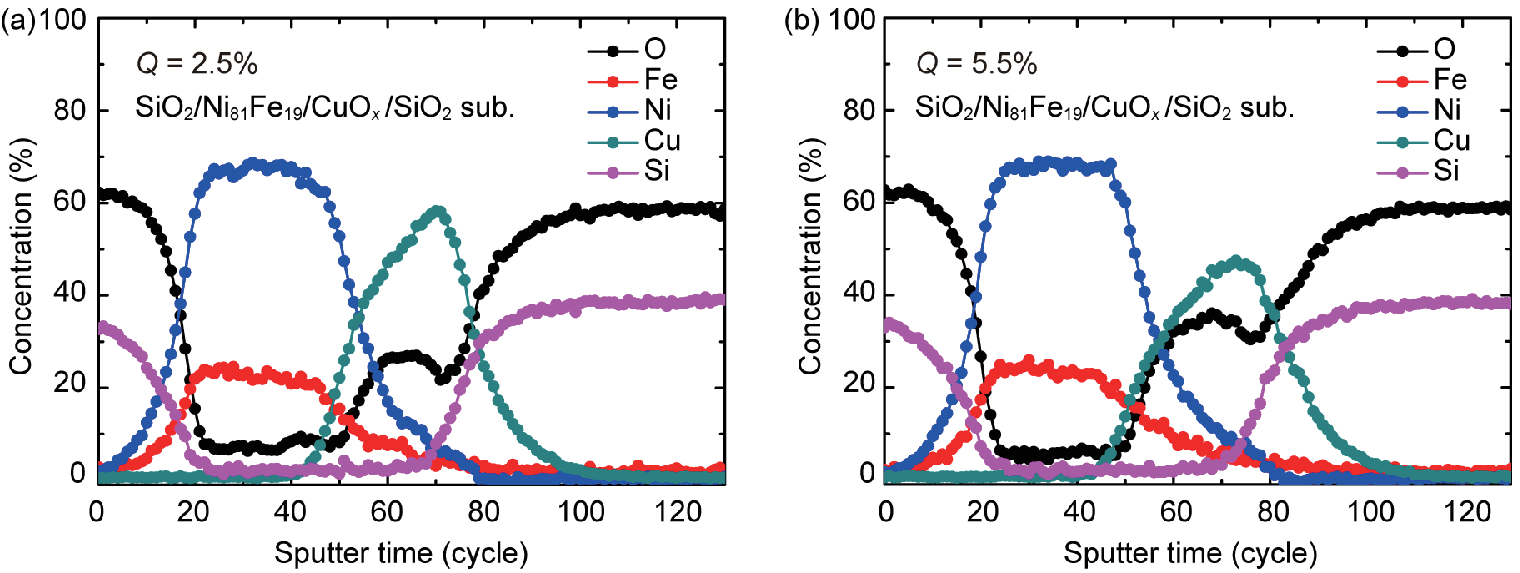}
\caption{The Auger electron spectroscopy (AES) depth profile of the Ni$_{81}$Fe$_{19}$(7.5 nm)/CuO$_x$(10 nm) bilayer film with (a) $Q=2.5\%$, and (b) $Q=5.5\%$.}
\label{figS4} 
\end{figure}

To discuss the possibility of current inhomogeneity resulted from the nonuniform oxidation of the Ni$_{81}$Fe$_{19}$ layer in the Ni$_{81}$Fe$_{19}$/CuO$_x$ bilayers, we performed Auger electron spectroscopy (AES) combined with the Ar-ion sputtering to detect the distribution depth profile of the elements. Figures~\ref{figS4}(a) and \ref{figS4}(b) show the AES concentration depth profiles of the Ni$_{81}$Fe$_{19}$(7.5 nm)/CuO$_x$(10 nm) films with $Q=2.5\% $ and $5.5\%$, respectively. The sputter time marked on the horizontal axis corresponds to the depth with the unit of sputtering circles. We found that the Ni$_{81}$Fe$_{19}$ layers are slightly oxidized with the concentration of O less than 7.4 at.\%, while the amount of O shows no obvious change between the cases of $Q=2.5\% $ and $5.5\%$, which is consistent with the fact that the resistance of the bilayers, around 1000 $\Omega$ for the microstrip used in ST-FMR measurements, shows no systematic dependence on $Q$. Moreover, it can be clearly seen that in both these two cases the Ni$_{81}$Fe$_{19}$/CuO$_x$ interfaces are well formed, and all the elements inside the Ni$_{81}$Fe$_{19}$ layers are distributed uniformly. Thus, we can conclude that in our Ni$_{81}$Fe$_{19}$(7.5 nm)/CuO$_x$(10 nm) bilayer structure, the element distribution of the Ni$_{81}$Fe$_{19}$ layer is quite uniform along the film normal direction. This shows that the Oersted field due to the possible current inhomogeneity caused by the nonuniform oxidation of the Ni$_{81}$Fe$_{19}$ layer is negligible in the Ni$_{81}$Fe$_{19}$/CuO$_x$ bilayers. We have also noticed that the atomic ratio of Ni to Fe obtained from AES measurements is~75.1 at.\%, which is smaller than the nominal one 81.0 at.\% used in this paper. For the CuO$_x$ layer, the depth profiles show that with the increase of Q values from $Q=2.5\%$ to $5.5\%$, the atomic ratio of O to Cu is significantly enhanced. These results are in good agreement with the XPS measurements.

\clearpage

\noindent  \textbf{4. Oersted field due to nonuniform current distribution in the Ni$_{81}$Fe$_{19}$ layer}
 \vspace{12pt}
 
\begin{figure}[hb]
\includegraphics[scale=1]{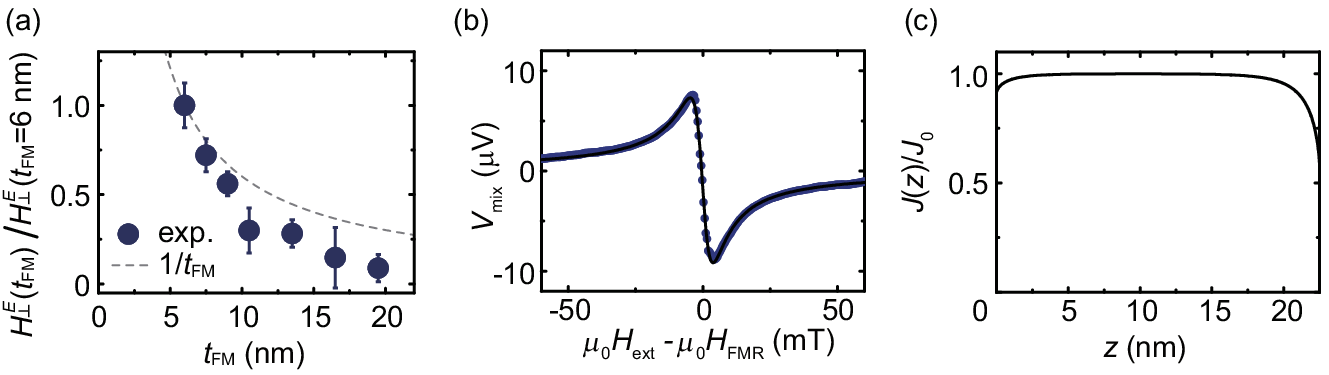}
\caption{(a) The Ni$_{81}$Fe$_{19}$-layer-thickness $t_\text{FM}$ dependence of $H_{\perp}^E=H_{\perp}/E$ for the Ni$_{81}$Fe$_{19}$($t_\text{FM}$)/CuO$_x$(10 nm) bilayer, where $H_{\perp}=H_\text{DL}+H_\text{Oe}^{\perp}$. The solid circles are the experimental data and the dashed curve is a function proportional to $1/t_\text{FM}$. (b) The $ H_\text{ext}$ dependence of the DC voltage $V_\text{mix}$ for the Ni$_{81}$Fe$_{19}$(22.5 nm)/CuO$_{x}$(10 nm) film with $Q=3.0$\% measured at 7 GHz. (c) The current-density $J(z)$ distribution in the Ni$_{81}$Fe$_{19}$ layer of the Ni$_{81}$Fe$_{19}$(22.5 nm)/CuO$_x$(10 nm) bilayer structure, where $z =0$ and $z =22.5$ nm correspond to the Ni$_{81}$Fe$_{19}$/CuO$_x$ and SiO$_2$/Ni$_{81}$Fe$_{19}$ interfaces, respectively. $J_0$ is the maximum current density in the film. The parameters of $p_1$, $p_2$, and $\lambda$ are 5/6, 1/6 and 2.7 nm, respectively, determined by Eqs.~(\ref{S4})-(\ref{S6}) based on the ST-FMR results.}
\label{figS5} 
\end{figure}

The design of our ST-FMR device is a rectangular microstrip of the Ni$_{81}$Fe$_{19}$/CuO$_x$ bilayer film connected to the Pt electrode, which has a ground-signal-ground structure and compatible with mirowave probe as shown in Fig.~1(a) of the main text. This high symmetrical arrangement is able to geometrically minimize the out-of-plane effective field ($H_\text{Oe}^{\perp}$) originated from Oersted field when the RF current flowing through the microstrip~\cite{PhysRevB.78.104401.1}. In Fig.~\ref{figS5}(a), we show Ni$_{81}$Fe$_{19}$-layer-thickness $t_\text{FM}$ dependence of the out-of-plane field $H_{\perp}$, where $H_{\perp}=H_\text{DL}+H_\text{Oe}^{\perp}$. Figure~\ref{figS5}(a) shows that $H_{\perp}$ is reduced to a negligible value with increasing the Ni$_{81}$Fe$_{19}$-layer thickness $t_\text{FM}$. This result indicates that the Oersted field $H_\text{Oe}^{\perp}$ contribution to $H_{\perp}$ is negligible in our device, since with increasing $t_\text{FM}$, the Oersted field $H_\text{Oe}^{\perp}$ due to nonuniformity should increase, while the DL-SOT effective field $H_\text{DL}$ should vanish. Because of $H_\text{Oe}^{\perp} \simeq 0$, we use $H_{\perp}\simeq  H_\text{DL}$ in the main text.

A nonzero in-plane effective field ($H_\text{Oe}^{\parallel}$) originated from Oersted field can still exist within our device, although the CuO$_x$ underlayer is semi-insulating. The origin of this in-plane Oersted field can be ascribed to the following two reasons: (i) a vertical current flow between the electrode and the sample due to the direct contact of the electrodes to the top surface of microstrip at its both ends and (ii) a nonuniform current distribution due to the different electron reflection coefficient from the top and bottom interfaces of the magnetic layer~\cite{10.1063/1.3006005.1}. However, in the Ni$_{81}$Fe$_{19}$/CuO$_x$ film, capped with the SiO$_2$ film, used in this study, the vertical current plays a minor role in the Oersted-field creation, since the top surface of the Ni$_{81}$Fe$_{19}$ layer is separated from the electrode by the SiO$_2$ cap layer. The Oersted field in the Ni$_{81}$Fe$_{19}$/CuO$_x$ film can be created only by the nonuniform current distribution within the Ni$_{81}$Fe$_{19}$ layer.

In thin metallic films, the nonuniform current distribution along the film normal direction ($z$) can be evaluated from the Fuchs-Sondheimer model~\cite{10.1080/00018735200101151.1,10.1063/1.1703100.1}, including a specularity parameter $p$ at each interface. The value of $p$ is in the range 0$\leq$$p$$\leq$1. In the case of $p=1$, the electron reflection is perfectly specular at the interface, while $p=0$ means that it is random, causing a reduced current. This reduction of the charge current extends over a length scale in $z$ away from interface, which is characterized by the mean free path $\lambda$. In the thick limit ($t_\text{FM}$$\gg$$\lambda$), the current distribution within the Ni$_{81}$Fe$_{19}$ layer is given by~\cite{10.1063/1.3006005.1}
\begin{equation}
J(z)=J_\text{0} \left[1-\frac{1-p_\text{1}}{2}f\left(\frac{z}{\lambda}\right)-\frac{1-p_\text{2}}{2}f\left(\frac{t_\text{FM}-z}{\lambda}\right)\right],
\tag{S4}
\label{S4} 
\end{equation}
where the function $f(x)$ is an exponential integral
\begin{equation}
f(x)=\frac{3}{2}\int^{1}_\text{0}\exp\left(-\frac{x}{u}\right)(1-u^2)du.
\tag{S5} \label{S5}
\end{equation} 
Here, $J_0$ is the maximum current density in the longitudinal direction within the Ni$_{81}$Fe$_{19}$ layer. $p_1$ and $p_2$ are the specularity parameters corresponding to the Ni$_{81}$Fe$_{19}$/CuO$_x$ and the SiO$_2$/Ni$_{81}$Fe$_{19}$ interfaces, respectively. The film normal is in the $z$ direction that defines the Ni$_{81}$Fe$_{19}$/CuO$_x$ interface at $z=0$ and the SiO$_2$/Ni$_{81}$Fe$_{19}$ interface at $z=t_\text{FM}$. Apparently, if the magnitude of $p_1$ and $p_2$ is different, it will give rise to an asymmetric current distribution within the Ni$_{81}$Fe$_{19}$ layer, which can generate the Oersted field $H_\text{Oe}^{\parallel}$. Considering the fact that the width of the microstrip ($w$) is much larger than the Ni$_{81}$Fe$_{19}$ thickness, i.e., $w$$\gg$$t_\text{FM}$, the averaged Oersted field due to the nonuniform current distribution is~\cite{10.1063/1.3006005.1}.
\begin{equation}
H_\text{Oe}^{\parallel}=-\frac{1}{t_\text{FM}}\int^{t_\text{FM}}_\text{0}J(z)\left(z-\frac{t_\text{FM}}{2}\right)dz.
\tag{S6} \label{S6}
\end{equation} 
Thus, Eqs.~(\ref{S4})-(\ref{S6}) enable us to estimate the upper bound of the Oersted field $H_\text{Oe}^{\parallel}$ in the Ni$_{81}$Fe$_{19}$/CuO$_x$ film. In order to extract the parameters of $p_1$, $p_2$, and $\lambda$, we performed the ST-FMR measurement on a Ni$_{81}$Fe$_{19}$(22.5 nm)/CuO$_x$(10 nm) bilayer film with $Q=3.0$\%, as shown in Fig.~\ref{figS5}(b). In this thick film, the FL effective field $H_\text{FL}$ due to the SOT vanishes, and only the Oersted field $H_\text{Oe}^{\parallel}$ contributes to the in-plane field: $H_{\parallel}=H_\text{FL}+H_\text{Oe}^{\parallel} \simeq H_\text{Oe}^{\parallel}$. Using Eq.~(2) of the main text, we deduce the effective field per unit current density $H_\text{Oe}^{\parallel}/J_\text{av}=(1.85\pm 0.04)\times 10^{-8}\text{ Oe/Acm}^{-2}$ for the Ni$_{81}$Fe$_{19}$(22.5 nm)/CuO$_x$(10 nm) film. Assuming that for a relative thick film the maximum current density ($J_0$) equals the average current density ($J_\text{av}$), namely $J_0 \simeq J_\text{av}$, we can reproduce the measured $H_\text{Oe}^{\parallel}/J_\text{av}$ for the $t_\text{FM}=22.5$ nm film, using Eqs.~(\ref{S4})-(\ref{S6}) with slightly tuned $\lambda=2.7$ nm~\cite{10.1063/1.3006005.1}, $p_1=5/6$, and $p_2=1/6$. We also notice that the Oersted field due to the nonuniform current distribution has the same positive sign as the $H_{\parallel}$ in the Ni$_{81}$Fe$_{19}$(7.5 nm)/CuO$_x$(10 nm) film.

Figure~\ref{figS5}(c) shows the current distribution in the Ni$_{81}$Fe$_{19}$ layer with the thickness of $t_\text{FM}=22.5$ nm determined using the above result. The current density at the Ni$_{81}$Fe$_{19}$/CuO$_x$ side is greater than that at the SiO$_2$/Ni$_{81}$Fe$_{19}$ side, as expected from the line shape of the ST-FMR signal shown in Fig.~\ref{figS5}(b). The ratio $J_\text{av}/J_0$ in Fig.~\ref{figS5}(c) is $\sim$0.98 in agreement with our assumption. Using the extracted parameters of $p_1$, $p_2$, and $\lambda$, the upper bound of the Oersted field $H_\text{Oe}^{\parallel}$ for the Ni$_{81}$Fe$_{19}$(7.5 nm)/CuO$_x$(10 nm) bilayer film can be estimated. Using the parameters determined for the Ni$_{81}$Fe$_{19}$/CuO$_x$ bilayer, we obtained $H_\text{Oe}^{\parallel}/J_\text{av}= 1.44\times 10^{-8}\text{ Oe/Acm}^{-2}$ for the $t_\text{FM}=7.5\text{ nm}$ film. This value is smaller than the in-plane effective field $H_{\parallel}/J_\text{av}=(1.85\pm 0.02)\times 10^{-8}\text{ Oe/Acm}^{-2}$ measured by the ST-FMR for the Ni$_{81}$Fe$_{19}$(7.5 nm)/CuO$_x$(10 nm) bilayer with $Q=3.0$\%. This indicates that the FL effective field $H_\text{FL}=H_{\parallel} -H_\text{Oe}^{\parallel}$ has a positive sign in the Ni$_{81}$Fe$_{19}$(7.5 nm)/CuO$_x$(10 nm) film with $Q=3.0$\%. Note that the magnitude of $H_{\parallel}/J_\text{av}$ is not exactly the same as that in the main text due to the irreversible changes on our sputtering system.

As shown in Fig.~3(a) of the main text, the measured $\xi_\parallel^E$, or $H_{\parallel}=H_\text{FL}+H_\text{Oe}^{\parallel}$, is negative at the high $Q$ values. We also note that the sign of $H_\text{Oe}^{\parallel}$ in the high $Q$ devices is the same as that in the device with $Q=3.0\%$, which has been confirmed by measuring the ST-FMR for a thick Ni$_{81}$Fe$_{19}$/CuO$_x$ film with $Q = 5.5\%$. This indicates that $H_\text{FL}$ opposes to the Oersted field $H_\text{Oe}^{\parallel}$ in the high $Q$ devices: $H_\text{FL}<0$ and $H_\text{Oe}^{\parallel}>0$. Since $H_\text{FL}>0$ at $Q=3.0$\% as discussed above, the sign of $H_\text{FL}$ is reversed from positive to negative by increasing the interfacial oxidation level in the Ni$_{81}$Fe$_{19}$(7.5 nm)/CuO$_x$(10 nm) bilayer.

\clearpage 
 
\noindent  \textbf{5. Field and frequency dependence of ST-FMR}
 \vspace{12pt}

\begin{figure}[hb]
\includegraphics[scale=1]{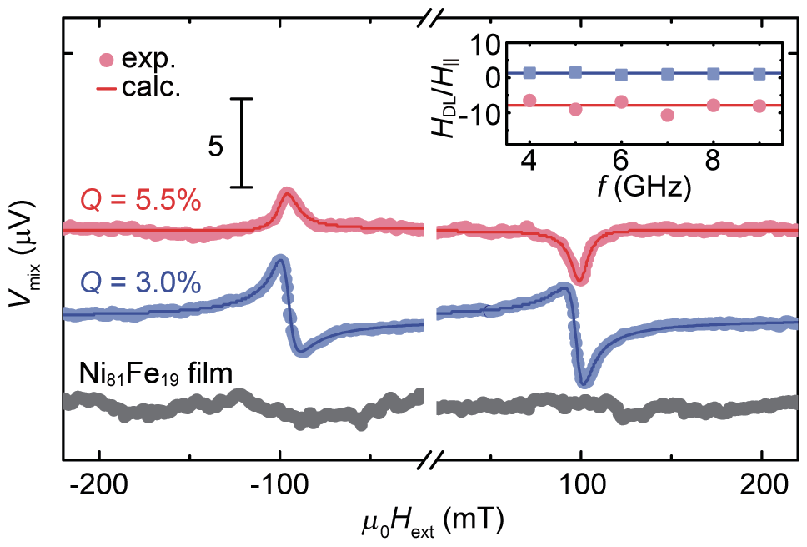}
\caption{The $ H_\text{ext}$ dependence of the DC voltage $V_\text{mix}$ for (a) the Ni$_{81}$Fe$_{19}$(7.5 nm)/CuO$_{x}$(10 nm) bilayer film with $Q=5.5$\% (red), (b) the Ni$_{81}$Fe$_{19}$(7.5 nm)/CuO$_{x}$(10 nm) bilayer film with $Q=3.0$\% (blue), and (c) the Ni$_{81}$Fe$_{19}$(7.5 nm) film (black), measured at the RF current frequency of $f=7$ GHz. The positive and negative applied magnetic field correspond to the magnetization oriented at $45^\circ$ and $225^\circ$ relative to the applied electric field, respectively. The inset shows $f$ dependence of $H_\text{DL}/H_{\parallel}$ for the Ni$_{81}$Fe$_{19}$(7.5 nm)/CuO$_{x}$(10 nm) bilayer film with $Q=5.5$\% (red) and $Q=3.0$\% (blue).}
\label{figS6} 
\end{figure}

We have performed the ST-FMR measurement on a SiO$_2$(4 nm)/Ni$_{81}$Fe$_{19}$(7.5 nm) film, fabricated on a SiO$_2$ substrate, using the same microstrip as that in the main manuscript. As shown in Fig.~\ref{figS6}, no ST-FMR signal is generated in the Ni$_{81}$Fe$_{19}$ single layer film. This result confirms that the current flow in the electrode of our device is very symmetrical, neither the imbalance current back-flow between the two ground contacts nor the imperfect centering of the microstrip creates a detectable ST-FMR signal.

In Fig.~\ref{figS6}, we also show the $ H_\text{ext}$ dependence of the DC voltage $V_\text{mix}$ for the Ni$_{81}$Fe$_{19}$(7.5 nm)/CuO$_{x}$(10 nm) bilayer films with $Q=5.5$\% and $Q=3.0$\%. This result shows that by reversing the external magnetic field direction, the sign of the voltage also changes correspondingly, as expected for the voltage generation induced by the ST-FMR.

In the Ni$_{81}$Fe$_{19}$/CuO$_{x}$ bilayers, the antisymmetric $A$ component in the ST-FMR signal corresponds to the in-plane effective field ($H_{\parallel}=H_\text{FL}+H_\text{Oe}^{\parallel}$) related to the FL effective field $H_\text{FL}$ and the Oersted field $H_\text{Oe}^{\parallel}$. In this situation, the $S/A$ ratio is directly related to $H_\text{DL}/H_{\parallel}$ as 
\begin{equation}
\frac{H_\text{DL}}{H_{\parallel}}=\frac{S}{A}\left( 1+\frac{\mu_0 M_\text{eff}}{\mu_0 H_\text{FMR}} \right)^{1/2},
\end{equation}
where $H_{\text{DL}}$ is the DL effective field. As shown in the inset to Fig.~\ref{figS6}, the $H_\text{DL}/H_{\parallel}$ ratio obtained from the ST-FMR shape is independent of the RF current frequency $f$, as expected.

\end{document}